\documentclass[a4paper]{article}
\usepackage{calc}
\usepackage{fullpage} 
\usepackage{pstricks}
\usepackage{multicol}
\usepackage{graphicx}
\usepackage{movie15}
\usepackage{hyperref}
\usepackage[version=3]{mhchem}
\usepackage{amsmath} 
\usepackage{amsfonts}
\usepackage{soul}
\usepackage[margin=1.7cm,top=3cm,bottom=2.8cm,centering]{geometry}
\usepackage{array}
\usepackage{array,multirow,makecell} 

\definecolor{rosepale}{rgb}{1.0, 0.7, 1.0}


\title{A role for ATP-dependent chromatin remodeling in the hierarchical cooperativity between noninteracting transcription factors}

\author{Denis Michel \\
\\
      \begin{small} Universite de Rennes1. Campus de Beaulieu Bat.13. 35042 Rennes France. E.mail: denis.michel@live.fr \end{small}\\}

\date{} 

\begin{document}
\maketitle

\begin{multicols}{2}
\noindent
\textbf{Abstract} \\
\newline
Chromatin remodeling machineries are abundant and diverse in eukaryotic cells. They have been involved in  a variety of situations such as histone exchange and DNA repair, but their importance in gene expression remains unclear. Although the influence of nucleosome position on the regulation of gene expression is generally envisioned under the quasi-equilibrium perspective, it is proposed that given the ATP-dependence of chromatin remodeling enzymes, certain mechanisms necessitate non-equilibrium treatments. Examination of the celebrated chromatin remodeling system of the mouse mammary tumor virus, in which the binding of transcription factors opens the way to other ones, reveals that breaking equilibrium offers a subtle mode of transcription factor cooperativity, avoids molecular trapping phenomena and allows to reconcile previously conflicting experimental data. This mechanism provides a control lever of promoter responsiveness to transcription factor combinations, challenging the classical view of the unilateral influence of pioneer on secondary transcription factors. \\


\section{Introduction}
The importance of ATP-dependent machineries remodeling chromatin by actively moving nucleosomes relatively to DNA, remains puzzling. Beside their possible structural roles in chromatin organization, nucleosome repositioning, histone exchange and DNA repair, a role in transcriptional cooperativity is proposed here.
The specificity and intensity of gene expression is governed by interactions between regulatory DNA sequences (cis-regulators) and various trans-acting factors (transcription factor proteins (TFs) and non-coding RNAs). The occupation of a gene promoter by these trans-regulators involves both micro-reversible and micro-irreversible steps. Micro-reversible binding processes can lead to sigmoidal concentration-dependent response through classical multimeric cooperativity (Bolouri and Davidson, 2002; Michel, 2010). The role of nucleosomes has also been examined from the micro-reversible perspective (Dodd et al., 2007; Segal and Widom, 2009; Mirny, 2010). The rapid equilibration of these thermally-driven phenomena, relatively to the slow changes of cellular components, simplifies the definition of the input functions used in gene network modeling (Bintu et al., 2005; Michel, 2010). But promoter occupancy also involves some micro-irreversible transitions such as chromatin remodeling and active dissociation processes. Precisely, it is proposed in the present study that inserting micro-irreversible steps in the process of promoter saturation, offers additional possibilities of potent cooperativity. A single example has been selected because it remarkably illustrates how micro-irreversible transitions can generate a refined discernment in gene expression. In this example, the micro-irreversible step corresponds to the energy-dependent phenomenon of chromatin-remodeling, in which the position of DNA around nucleosomes is modified. By this way, upon binding to DNA, a first TF directs the accessibility to DNA of other ones. This mechanical activity allows: (i) cooperativity between non-physically interacting TFs and (ii) constitutively expressed TFs to participate to conditional induction. One of the most celebrated chromatin remodeling system is provided by the thoroughly documented Mouse Mammary Tumor Virus (MMTV) promoter. 

\section{Breaking hierarchical polymerization is necessary to maintain molecular dynamics}

The assembly of macromolecular complexes generally proceeds in a hierarchical manner in the cell. For example, a component $ C $ which cannot bind to the isolated components $ A $ and $ B $, can bind to a pre-associated complex $ AB $. Hierarchical binding chains such as $ A + B \rightleftharpoons  AB, + C \rightleftharpoons  ABC, + D \rightleftharpoons  ABCD $… etc, are often involved in the building of multi-molecular complexes, but are less compatible with the dynamic and reactive behaviours of solubles components. Indeed, in equilibrium conditions, the chain written above leads to the trapping of the early components in the complexes as long as the late components are present. This phenomenon can exist for TF binding to gene promoters. It can hold for example, for the successive binding steps observed in equilibrium conditions between the TodT TFs and the series of TodT-binding sites juxtaposed in the Tod gene promoter, that has been proposed to be mediated by DNA conformation changes (Lacal et al., 2008). Beside this puzzling situation of equilibrium allostery, the hierarchical binding of transcription factors in equilibrium conditions is also possible in the case of the large eukaryotic preinitiation complexes made of the so-called general transcription factors GTFs (Michel, 2010). But hierarchical relationships have also been reported for non-interacting isolated TFs in absence of any trapping phenomenon. To allow reconciling hierarchical binding and absence of trapping, one should postulate the possibility to break equilibrium. This situation is well illustrated by the case of the occupation of the MMTV promoter involving micro-irreversible processes, thoroughly documented but not yet clearly understood. In section 4, this system will be analysed under the classical equilibrium assumption. Then, in section 5, a non-equilibrium scheme will be proposed based on hypotheses built from MMTV experimental data, in which the equilibrium-breaking machines are the ATP-dependent SWI/SNF chromatin remodeling enzymes.

\section{Data obtained for the MMTV promoter occupancy are irreconcilable from the time-reversible perspective}
A central piece of data about MMTV expression is the role of nucleosomes in the mutual influence between the glucocorticoid receptor (GR) and a group of TFs (NF1/Oct). Although the activation of MMTV by GR and NF1/Oct-1 seemed clear in the initial reports, discrepancies appeared in following studies. The basis of glucocorticoid hormone-induced MMTV regulation is that GR has an initiating role, triggered upon hormone binding (stress hormone corticol or corticosterone) and subsequently amplified by NF1 and Oct-1. This sequential action is dependent on the position of nucleosomes on DNA, since it is not observed with naked DNA (Richard-Foy and Hager, 1987; Archer et al., 1992; Ch\'avez and Beato, 1997). The repositioning of nucleosomes triggered by GR, probed by nuclease or chemical mapping, leads to the exposure of the NF1 and Oct-1 binding sites and is mediated by SWI-SNF ATPases (Fryer and Archer, 1998). The different roles of GR and NF1 in initiating and amplifying transcription respectively, are explained by their differential mode of interaction with chromatin:  GR can bind to DNA wrapped around nucleosomes, contrary to NF1 which requires a fully accessible double helix (Eisfeld et al., 1997). This DNA-binding hierarchy, first of hormone-bound GR and then of NF1/Oct, offers a powerful opportunity of cooperativity. Indeed, GR is inducible but not very potent contrary to the couple NF1/Oct. As NF1/Oct can access DNA only upon binding of GR, the promoter activity can become strongly sigmoidal, particularly if NF1/Oct is transcriptionally more potent than GR. Sigmoidal responses are generally due to decreased responsiveness to low signals. This is the case for the MMTV promoter in which TFs are prevented to bind at low concentration. But this elegant mechanism has then been clouded in the following reports, which introduced new actors and revised the hierarchy of binding of GR and NF1. In sharp contrast with the earlier articles, NF1 and Oct-1 binding sites have been shown to preset chromatin prior to GR binding (Belikov et al., 2004). The picture is thus more complex than supposed previously and the mutual influence between GR and NF1 for binding to the MMTV promoter is now described as dualistic (Hebbar and Archer, 2007), blurring the logic of this system. The same apparent paradox has been pointed for the relationships between purported "pioneer" and secondary transcription factors (Caizzi et al., 2014). In fact, the strict dependence on the previous fixation of a factor to allow the fixation of another factor can be alleviated if introducing an additional step of chromatin remodeling (Fig.1). It will be shown that the cooperative relationships between GR and NF1, which are unclear when examined only from a time-reversible perspective, can be usefully reconsidered from a non-equilibrium perspective (section 5), but the outcomes of equilibrium modeling is first examined below for comparison.

\section{Equilibrium modeling of hierarchical MMTV promoter occupancy}
In the simplest hierarchical modeling scheme assuming micro-reversibility (Fig.1a), MMTV transcription is stimulated by two groups of transcription factors GR (named $ A $) and NF1/Oct (named $ B $) (Fig.1). $ A $ and $ B $ bind to the MMTV promoter ($ P $) through their DNA-binding domain (DBD) in a hierarchical manner, but once bound to DNA, they are supposed to stimulate transcription in an independent and additive manner, through their activation domain (TAD). In these conditions, the fractional activity ($ F $) ranging from 0 to 1, of the MMTV promoter, is described in Eq.(1). 

\begin{equation} F=\frac{p(A)\text{ }k_{A}+p(B)\text{ }k_{B}}{k_{A}+k_{B}} \end{equation}

In this equation, $ k_{A} $ and $ k_{B} $ are the maximal frequencies at which $ A $ and $ B $, when bound to DNA, recruit transcription machineries, thereby initiating multiple rounds of transcription. These frequencies should be weighted by the probabilities of presence of $ A $ and $ B $ on the promoter, written $ p(A) $ and $ p(B) $ respectively (with small letters $ p $ to not be confused with the promoter $ P $).

\begin{center}
\includegraphics[width=7cm]{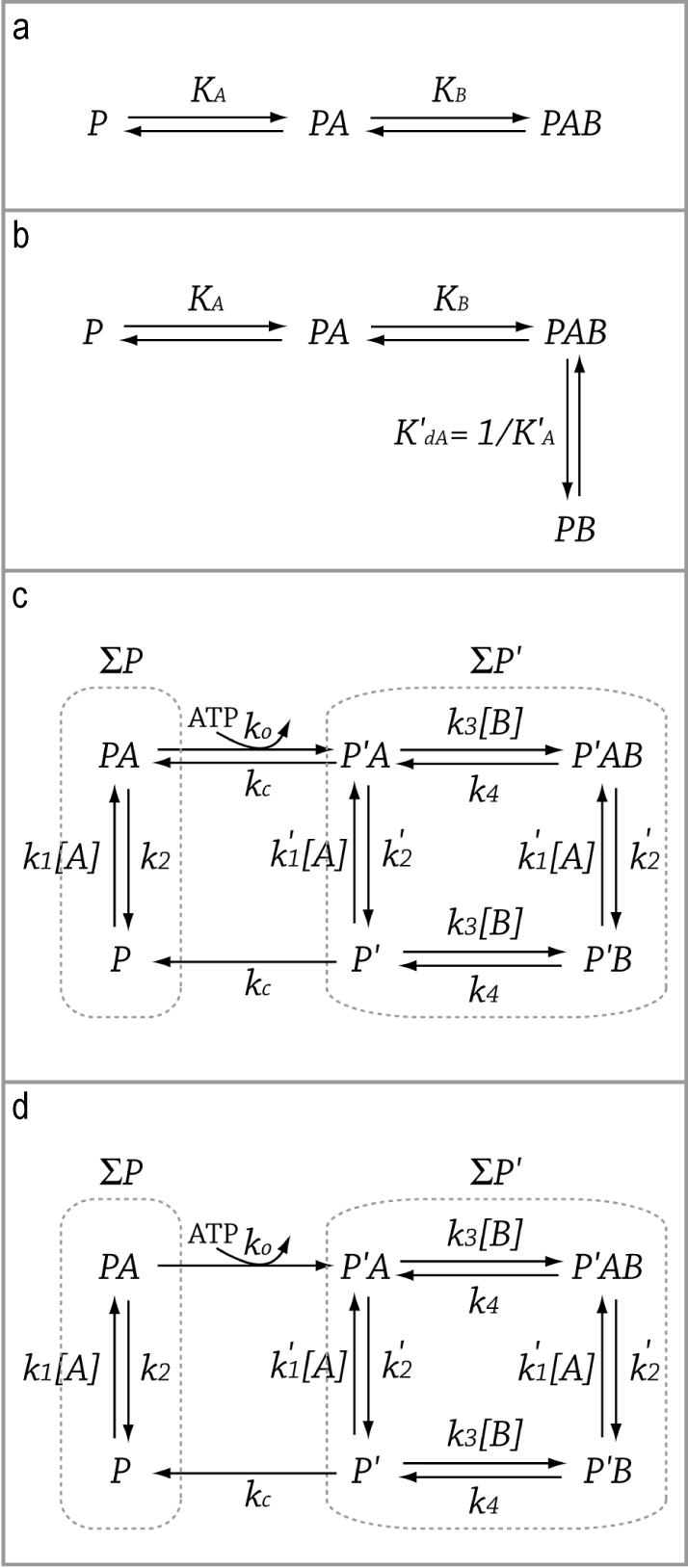} \\
\end{center}
\begin{small} \textbf{Figure 1.} Different models to explain the hierarchical occupation of the MMTV promoter ($ P $) by GR (named $ A $) and NF1 (named $ B $). The schemes (\textbf{a}) and (\textbf{b}), comply with the principle of microscopic reversibility but not the schemes (\textbf{c}) and (\textbf{d}). In (\textbf{c}), $ k_{o} $ is the rate of chromatin opening driven by SWI/SNF ATPases and $ k_{c} $ is the rate of chromatin closing, driven by stabilization of DNA bending. $ B $ cannot bind to $  P $ because of inappropriate nucleosome positioning, while $ A $ can bind to both $ P $ and $ P' $ with different constants. In the scheme (\textbf{d}), chromatin closing can occur only when the promoter is free of any TF. \end{small}\\

The probabilities $ p(A) $ and $ p(B) $, equivalent to fractional occupation times, can be formulated through an Adair approach, as the ratio of occupied over total binding sites, which can be expressed as concentrations in ergodic conditions, by enumerating the possible promoter states. 

\begin{subequations} \label{E:gp}  
\begin{equation} p(A)=\dfrac{[PA]+[PAB]}{[P_{0}]+[PA]+[PAB]}\end{equation} \label{E:gp1}
\noindent
and, given that $ B $ is supposed to access DNA only when $ A $ is present,
\begin{equation} p(B)=\dfrac{[PAB]}{[P_{0}]+[PA]+[PAB]} \end{equation} \label{E:gp2}
\end{subequations} 

Using $ P_{0} $ as a reference, Eq.(2) can be converted into

\begin{subequations} \label{E:gp}  
\begin{equation} p(A)=K_{A}[A](1+K_{B}[B])/D \end{equation} \label{E:gp1}
\begin{equation} p(B)=K_{A}[A]K_{B}[B]/D \end{equation} \label{E:gp2}
with
\begin{equation} D=1+K_{A}[A]+K_{A}[A]K_{B}[B] \end{equation} \label{E:gp3}
\end{subequations} 

In this scheme, GR is prevented to dissociate from a saturated promoter. This trapping effect which can appear puzzling, is inherent to the equilibrium modeling of sequential cooperativity, but such a trapping of GR on the MMTV promoter is not consistent with the observation that GR can escape DNA whatever the chromatin configuration (Fletcher et al., 2000; Hager et al., 2000). To avoid this problem, one can imagine an alternative scenario (Fig.1b), in which GR dissociation does not require the absence of NF1. Two different equilibrium constants are defined for GR ($ K_{A} $ and $ K'_{A} $ = 1/$ K'_{dA} $), to take into account the different chromatin states. The occupation probabilities of the $ A $ and $ B $ sites are respectively:

\begin{subequations} \label{E:gp}  
\begin{equation} p(A)=K_{A}[A](1+K_{B}[B])/D \end{equation} \label{E:gp1}
\begin{equation} p(B)=K_{A}[A]K_{B}[B](1+K'_{dA}/[A])/D \end{equation} \label{E:gp2}
with
\begin{equation} D=1+K_{A}[A](1+K_{B}[B](1+K'_{dA}/[A])) \end{equation} \label{E:gp3}
\end{subequations} 

But the phenomenon of trapping now concerns NF1, possibly for long periods in case of removal of the glucocorticoid hormone. Though puzzling, this possibility would be consistent with the observation that NF1 is present on DNA prior to hormone addition (Hebbar and Archer, 2003; Belikov et al., 2004). But it remains to explain why the MMTV promoter would be inactive in spite of the continuous presence of NF1. A possible explanation could be that NF1 doesn't work as long as it is prevented to recycle on DNA, according to the model of one-shot TFs like ATF6 (Michel, 2010). This system would still conform microscopic reversibility, but the notion of equilibrium would become shaky since it is no longer dynamic after disappearance of $ A $. To propose a more plausible formulation of the MMTV promoter occupancy, relieved from any trapping effect, a micro-irreversible steps should be introduced in the system.

\section{Non-equilibrium modeling of the MMTV promoter occupancy}
 
To not recourse to trapping phenomena which are not experimentally verified for MMTV, one should postulate another mode of cooperativity, liberated from the micro-reversibility constraints. Energy inputs obviously exist in the system and are provided by ATPases (SWI/SNF), recruited by DNA-bound GR (Fryer and Archer, 1998). Among the different micro-irreversible mechanisms that can be imagined, the model shown in Fig.1c is an attempt to reconcile the more experimental data as possible, in a novel scheme as simple as possible. Chromatin remodeling can be triggered and reversed dynamically, according to the well established reversibility (in its traditional acceptation) of hormone-induced nucleosome positioning (Belikov et al., 2001) and to the dynamic interaction of remodeling complexes with the MMTV promoter (Johnson et al., 2008). The spontaneous nucleosome repositioning from $ P' $ to $ P $ is dictated by the intrinsic bendability of DNA sequence and can be considered as nearly micro-irreversible. Since transient behaviours following GR activation can be neglected for the resulting gene expression, a steady state treatment is sufficient for modeling this system. A reasonable additional hypothesis is that the time scales are different between the rapid equilibration of the TFs with the promoter, and the slower dynamics of micro-irreversible chromatin remodeling, for opening site for B ($ k_{o} $) and for closing it ($ k_{c} $). The time scale separation hypothesis is not always applicable, but is justified in the present case, given the rapid equilibration of the TFs with the MMTV DNA, suggested by the short turnovers of GR (12 milliseconds) evidenced by fluorescence recovery after photobleaching (FRAP)(Sprague et al., 2004). This condition allows to use the approach of (Cha, 1968), mixing in the same treatment rate constants (time-dependent) and equilibrium constants (time-independent). In this method, several groups of rapidly equilibrated species are defined using equilibrium constants. They correspond in the present case to the two chromatin states of the MMTV promoter, which will be written $ \Sigma P $ and $ \Sigma P' $ (Fig. 1c,d). Two DNA-binding constants $ K_{a} $ and $ K'_{a} $ are postulated for $ A $ depending on the chromatin state, but $ K'_{a} $ is not affected by the presence or not of $ B $, given that these TFs do not directly interact with each other.

\begin{subequations} \label{E:gp}  
\begin{equation} [\Sigma P]=[P_{0}]+[PA] \end{equation} \label{E:gp1}
\begin{equation} [\Sigma P']=[P'_{0}]+[P'A]+[P'B]+[P'AB] \end{equation} \label{E:gp2}
\end{subequations} 

with

\begin{subequations} \label{E:gp}  
\begin{equation} \frac{[P_{0}]}{[\Sigma P]}=\frac{1}{D_{P}} \end{equation} \label{E:gp1}
\begin{equation} \frac{[PA]}{[\Sigma P]}=\frac{K_{A}[A]}{D_{P}} \end{equation} \label{E:gp2}
and
\begin{equation} D_{P}=1+K_{A}[A] \end{equation} \label{E:gp3}
\end{subequations} 

For the $ P' $ states:

\begin{subequations} \label{E:gp}  
\begin{equation} \frac{[P'_{0}]}{[\Sigma P']}=\frac{1}{D_{P'}} \end{equation} \label{E:gp1}
\begin{equation} \frac{[P'A]}{[\Sigma P']}=\frac{K'_{A}[A]}{D_{P'}} \end{equation} \label{E:gp2}
\begin{equation} \frac{[P'B]}{[\Sigma P']}=\frac{K_{B}[B]}{D_{P'}} \end{equation} \label{E:gp3}
\begin{equation} \frac{[P'AB]}{[\Sigma P']}=\frac{K'_{A}[A]K_{B}[B]}{D_{P'}} \end{equation} \label{E:gp4}
with
\begin{equation} D_{P'}=(1+K'_{A}[A])(1+K_{B}[B]) \end{equation} \label{E:gp5}
\end{subequations} 

When gathering the promoter states, the probabilities $ p(A) $ and $ p(A) $ that $ P $ is occupied by $ A $ and $ B $ respectively, can be defined as follows

\begin{subequations} \label{E:gp}  
\begin{equation} p(A)=p(P\cap A)+p(P'\cap A) \end{equation} \label{E:gp1}
equivalent to
\begin{equation} p(A)=p(A|P)\text{ }p(P)+p(A| P')\text{ }p(P') \end{equation} \label{E:gp2}
and
\begin{equation} p(B)=p(P'\cap B) \ (\textup{given that} \ p(P\cap B)=0) \end{equation} \label{E:gp3}
\begin{equation} p(B)=p(B\cap P')\text{ }p(P') \end{equation} \label{E:gp4}
where
\begin{equation} p(A| P)= \dfrac{[PA]}{[\Sigma P]}=\dfrac{K_{A}[A]}{1+K_{A}[A]} \end{equation} \label{E:gp5}
\begin{equation} p(A| P')= \dfrac{[P'A]+[P'AB]}{[\Sigma P']} =\dfrac{K'_{A}[A]}{1+K'_{A}[A]} \end{equation} \label{E:gp6}
\begin{equation} p(B| P')=\dfrac{[P'B]+[P'AB]}{[\Sigma P']} =\dfrac{K_{B}[B]}{1+K_{B}[B]} \end{equation} \label{E:gp7}
and
\begin{equation} p(P)= \dfrac{[P]}{[P]+[P']} \ \textup{and} \ p(P')= 1-p(P) \end{equation} \label{E:gp8}
\end{subequations} 

$ [P] $ and $ [P'] $ are the amounts of time in which the promoter is in the $ P $ and $ P' $ states, which can be deduced from the steady state balance. If supposing, to agree with experimental observations, that the restoration of the basal chromatin can occur only when the promoter is not occupied by $ B $ (Fig.1c), then, the steady state can be written:

\begin{equation} k_{0} [P]\text{ }p(A| P)= k_{c} [P'] \left (1-p(B| P')  \right ) \end{equation}

which yields, using the values defined previously, 

\begin{equation} \frac{[P]}{[P']}=\frac{k_{c}(1+K_{A}[A])}{k_{o}K_{A}[A](1+K_{B}[B])} \end{equation}

leading to

\begin{subequations} \label{E:gp}  
\begin{equation} p(A)=K_{A}[A]\left (k_{c}+k_{o}K'_{A}[A]\left (\frac{1+K_{B}[B]}{1+K'_{A}[A]} \right )\right )/D \end{equation} \label{E:gp1}
and
\begin{equation} p(B)=k_{o}K_{A}[A]K_{B}[B]/D \end{equation} \label{E:gp2}
where
\begin{equation} D=k_{c}(1+K_{A}[A])+k_{o}K_{A}[A](1+K_{B}[B]) \end{equation} \label{E:gp3}
\end{subequations} 

These results are then incorporated in Eq.(1). The capacity of this system to generate sigmoidal curves is due to the products of the concentrations of $ A $ with itself (in Eq.(11a)) and with $ B $ (in Eqs(11a) and (11b)) if assuming a double time scale separation: (i) between DNA/TF interactions and chromatin remodeling kinetics, as previously postulated and (ii) between chromatin remodeling and gene product concentration changes.

\section{TF concentration-dependence of the promoter activity}
 
 The important parameters to evaluate are the sigmoidicity and sensitivity of the promoter activity to TF concentration changes. Sigmoidicity is classically obtained when the TFs should multimerize for binding to DNA. This is the case for GR which is active as a dimer. But to focus on the specific source of cooperativity provided by the mechanism examined here, dimerisation will be ignored and the TFs will be considered as preformed dimers. For easier analyses, the fractional promoter activity equations will be adimensioned by setting some constants. The ratio of the transcriptional strength of DNA-bound $ B $ and $ A $ is $ \gamma =k_{B}/k_{A} $. In the mechanisms of Fig.1c and 1d, the ratio of equilibrium constants of GR binding to the two chromatin states is $ \alpha =K_{A}/K'_{A} $, and the ratio of chomatin opening and closing rates is $ \beta =k_{o}/k_{c} $. $ \gamma $ is not an equilibrium constant and is modifiable but the cellular contents in remodeling enzymes and ATP. An energy-independent conformational equilibrium between $ PA $ and $ P'A $ would lead again to molecular trapping (of $ A $ and of the chromatin-remodeling enzyme). The equivalence between $ k_{o} $ and a Poissonian rate is a gross approximation since the transition $ PA \rightarrow P'A $ encloses many elementary events, including the recruitment of enzymes, of ATP, catalytic steps etc, which will not be detailed here.
	Let be $ x $ and $ y $ the binding potentials of $ A $ and $ B $ respectively which are, for the non-equilibrium mechanisms $ x=K'_{A}[A] $, $ y=K_{B}[B] $ and for the equilibrium mechanisms $ x=K_{A}[A] $ and $ y=K_{B}[B] $. Fractional activity can be defined with these symbols and used for drawing 3D plots. They are listed below for the different models.

\subsection{Independent binding of $ A $ and $ B $ to the promoter}
Eq.(1) yields

\begin{equation} F=\frac{1}{1+\gamma }\left (\frac{x}{1+x}+\gamma \frac{y}{1+y}  \right ) \end{equation}

The corresponding curve is shown in Fig.2a, when saturating $ A $ can trigger only one quarter of maximal activation ($ \gamma $= 3).

\subsection{Putative equilibrium hierarchical model of Fig.1a (Eq.(3))}

\begin{equation} F=\frac{x(1+y(1+\gamma ))}{(1+\gamma )(1+x+xy)} \end{equation}

The corresponding plot is shown in Fig.2b for $ \gamma $= 3.

\subsection{Putative equilibrium hierarchical model of Fig.1b (Eq.(4)).}

\begin{equation} F=\frac{x(1+y)+\gamma y(\alpha +x)}{(1+\gamma )(1+x+y(\alpha +x))} \end{equation}

\subsection{Non-equilibrium hierarchical model of Fig.1c (Eq.(11)).}
In this system, chromatin relaxation can occur during the periods of absence of $ B $, irrespective of whether $ A $ is present or not.
\begin{equation} F=\frac{\alpha x\left [ 1+\beta x \left (\frac{1+y}{1+x}  \right )+ \beta \gamma y\right ]}{(1+\gamma )[1+\alpha x (1+\beta (1+y))]} \end{equation}

\end{multicols}
\begin{center}
\includegraphics[width=13cm]{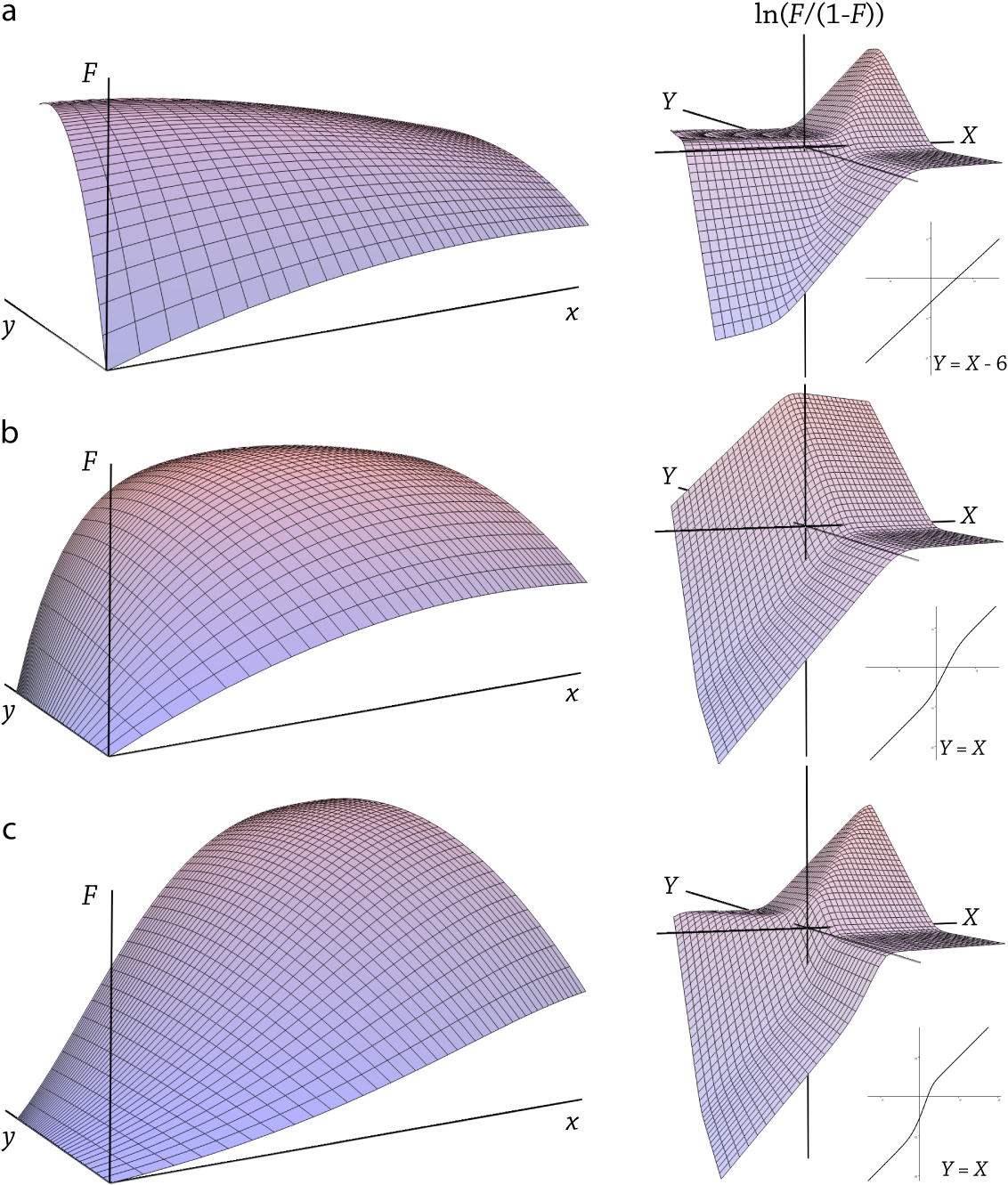} \\
\end{center}
\begin{small} \textbf{Figure 2.} Comparative shapes of promoter activity curves in linear coordinates (left panels) and in Hill coordinates (right panels). $ x $ and $  y $ are the binding potentials of the TFs $ A $ and $ B $ used in the main text, and $ X $ and $ Y $ are their logarithms. The small inserts show 2D sections of the Hill plots at the indicated planes. In all cases, $ B $ is considered 3-time more potent than $ A $ for activating transcription. (\textbf{a}) The two TFs bind independently to the gene promoter (Eq.(12)). (\textbf{b}) Putative equilibrium hierarchical cooperativity model (Eq.(13)). (\textbf{c}) Hierarchical model with chromatin remodeling, in which the basal chromatin organization state can be restored only from the TF-free promoter (Eq.(18)), with the combination of parameters ($ \alpha $, $ \beta $, $ \gamma $) = (0.001, 2, 3). $ A $ and $ B $ are assumed to participate to the recruitment of transcription machineries in an additive manner. \end{small}\\
\begin{multicols}{2}

\subsection{Non-equilibrium hierarchical model of Fig.1d.}

In this alternative possibility, chromatin closing to $ B $ can occur only for a TF-free promoter (specifically not from the $ P'A $ state). This possibility could for example be explained by the persistent molecular association between $ A $ and chromatin-remodeling enzymes. In this case, the same development as in section 5, gives:

\begin{equation} k_{o} [P]\text{ }p(A| P)= k_{c} [P'](1-p(A| P'))(1-p(B| P')) \end{equation}

leading to the following steady state $ P/P' $ ratio

\begin{equation} \frac{[P]}{[P']}=\frac{k_{c}(1+K_{A}[A])}{k_{o}K_{A}[A](1+K'_{A}[A])(1+K_{B}[B])} \end{equation}

and to the fractional activity 

\begin{equation} F=\frac{\alpha x [1+\beta x(1+y)+\beta \gamma  y(1+x)]}{(1+\gamma )[1+\alpha x (1+\beta (1+x)(1+y))]} \end{equation}

A representative plot of this condition is shown in Fig.2c for ($ \alpha $, $ \beta $, $ \gamma $) = (0.05, 2, 3). This set of parameters is chosen to agree with experimental observations. Indeed, $  B $ is considered as more potent than $ A $ because it is in fact not a single TF, but a combination of several potent TFs (NF1 and Oct). The higher affinity of $ A $ (GR) for the promoter ($ K'_{a} > K_{a} $), is suggested by the fact that the chromatin configuration permissive to NF1/Oct binding, strongly favours GR binding (Belikov et al., 2004).

\section{Functional opportunities offered by the system}
\subsection{A source of nonlinearity for structuring gene networks}

Sigmoidal genetic responses are the typical ingredients of multistable dynamic gene regularory networks. In a simple example, if a gene subject to this mode of transcriptional regulation stimulates its own expression through a positive feedback, then, the sigmoidal and saturable expression curve crosses twice the non-saturable first-order degradation line, thus generating bistability (Cherry and Adler, 2000). This general role of nonlinear responses in the formation of Boolean-like networks is not specific to hierarchical cooperativity and will not be detailed further here.

\subsection{Cooperativity between non-interacting TFs}
 The mechanism of transcriptional cooperativity most widely reported and modeled in the literature is that obtained with TFs capable of physically interacting with each other and binding to a series of non-consensual DNA elements in a promoter. By this way, the direct interactions between the TFs help them to bind together to DNA, whereas their individual affinity for their DNA elements would not have allowed their independent binding. But a recent study suggested that this mode of cooperativity is in fact doubtful (Chu et al., 2009), since direct interactions between adjacent TFs would lead to the clustering of TFs on DNA which inevitably contains non-specific TF binding sites. Indeed, all the TFs have a minimal non-specific affinity for DNA (at least electrostatic). This point is interesting since it suggests that many interactions experimentally shown between TFs to explain cooperativity, could result from experimental drawbacks in detecting protein interactions (Mackay et al., 2007). This problem no longer holds for the model of hierarchical cooperativity described here. Moreover, the number of chromatin-remodeling machineries in the cell (Rippe et al., 2007), which is so far intriguing, further supports the general importance of the present proposal.
 
\subsection{Participation of constitutively expressed TFs to conditional expression}

The NF1 and Oct-1 TFs are generally expressed at high level by the laboratory cell lines used in the MMTV experiments cited above. However, in spite of their constitutive presence, the expression of MMTV integrated in ordered chromatin, is very low in absence of glucocorticoid hormone (personal data not shown). Hence, NF1 and Oct-1 contribute to the MMTV transcriptional strength but not to the decision to transcribe or not to transcribe. 
 
\subsection{Adjustment of the degree of sensitivity and responsiveness of the promoter}

Expectedly, when chromatin remodeling is inhibited ($ \beta =0 $), Eqs.(15) and (18) reduce to a simple hyperbola $ \alpha x/(1+ \alpha x) $, but with chromatin remodeling, the behaviour of this system is unusual compared to classical modes of cooperativity. It can generate an "interrupter-like" mode of promoter functioning with strong non-linearity. For non-zero $ \alpha $ and $ \beta x $, Eq.(18) can approach the Hill-like equation $ \alpha \beta (1+\gamma)x^{2}y/(1+ \alpha \beta (1+\gamma)x^{2}y) $, where the square exponent of $ x $ in absence of any postulated dimerisation, reflects its dual participation in regulating both the $ P $ and $ P' $ promoter states. Interestingly, the self-cooperativity of $ x $ can be obtained even when $ B $ is transcriptionally inactive ($ \gamma = 0 $). In classical equilibrium mechanisms of promoter occupation, the responsiveness and sensitivity to increasing concentrations of activated TFs, is fixed by the physico-chemical cellular conditions and by the values of equilibrium constants, which are themselves dictated by macromolecule structures. For example in the case of TF dimerisation, the degree of cooperativity is not modifiable and determined by the affinity constants for given DNA elements. By contrast, in the present system, further adjustments are possible by tuning the activity and quantity of SWI/SNF chromatin remodeling enzymes (Fig.3) and the energy status of the cell (ATP), which influence altogether the $ \beta $ parameter. In particular, at given TF binding potentials, the chromatin-remodeling rate provides a precise control lever of the degree of promoter activation which can be coupled to additional threshold effects such as the buffering of low levels of mRNA expression by small RNAs, as shown in bacteria (Levine and Hwa, 2008). \\

\begin{center}
\includegraphics[width=7cm]{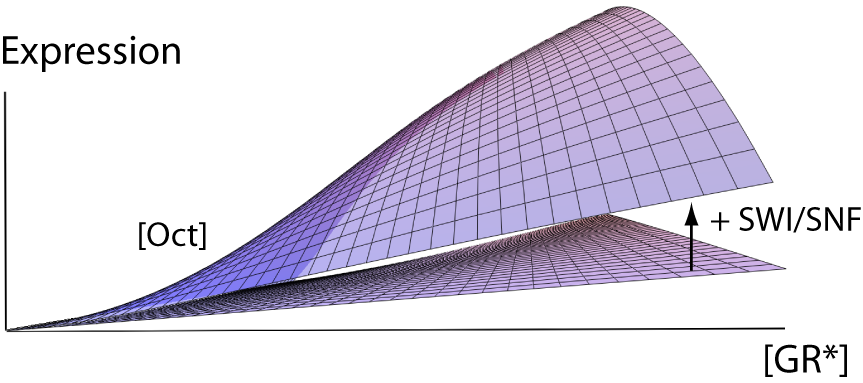} \\
\end{center}
\begin{small} \textbf{Figure 3.} Alternative view of the 3D plot of Fig.2c, showing the sigmoidal response of MMTV expression to the couple of activated TFs GR and Oct. For a given level of activity of SWI/SNF, this shape of the transcriptional activity surface makes the response of MMTV to GR and Oct strongly overadditive. Its near-horizontal slope at low GR and Oct values, allows to avoid inappropriate activation. Moreover, MMTV expression can be further adjusted by regulating the amount or the activity of the SWI/SNF enzymes and by other threshold effects which can cancel transcription activation by GR alone. \end{small}\\

Thought they are functionally important, the subtle differences between sigmoidal responses cannot be easily evaluated by eye in linear coordinates. Among the different mathematical tools developed for analysing non-hyperbolic fractional curves, the Hill representation long proved very useful because it allows to focus on the specificities of the systems (Cornish-Bowden and Koshland, 1975; Dahlquist, 1978). Using the logarithm of TF binding potentials allows to give to the range of binding potential between 0 to 1, the same importance than between 1 and infinite, for better visualizing the effects of ligand concentrations below midsaturation (for $ \ln(K[TF]) = 0 $). The logarithm of the ratio of fractional activity vs inactivity $ \ln(F/(1-F)) $ ("logit" coordinate), allows to finely appreciate the behaviour of the system, including the degree of cooperativity between the TF(s) through the slope of the curve. An hyperbolic phenomenon gives a slope of 1. Indeed, when $ F = x/(1+x) $, then $ F/(1-F) = x $, thus eliminating saturation effects. In addition, in multidimensional Hill plots (3D in the present case), the relative participation of the different actors in the course of saturation can be visualized. Although the Hill plots are generally used for equilibrium phenomena such as hemoglobin oxygenation, they can also apply to steady states. The Hill equations corresponding to Eqs.(15) and (18) are Eqs.(19) and (20) respectively:

\begin{equation} H_{(X,Y)}=\ln\frac{\alpha \textup{e}^{X}\left [1+\beta \textup{e}^{X} \left (\frac{1+\textup{e}^{Y}}{1+\textup{e}^{X}}\right )+\beta \gamma \textup{e}^{Y}\right ]}{1+\gamma +\alpha \textup{e}^{X}\left [1+\beta \left (\frac{1+\textup{e}^{Y}}{1+\textup{e}^{X}}\right )+\beta \gamma \right ]} \end{equation}

\begin{equation} 
H_{(X,Y)}=\ln\frac{\alpha \textup{e}^{X}[1+\beta \textup{e}^{X}(1+\textup{e}^{Y})+\beta \gamma \textup{e}^{Y}(1+\textup{e}^{X})]}{1+\gamma +\alpha \textup{e}^{X}[\gamma +\beta (1+\textup{e}^{Y}) +\beta \gamma (1+\textup{e}^{X})]} \end{equation}

where $ X = \ln(x) $ and $ Y = \ln(y) $. The corresponding plots are shown in the right panels of Fig.2 using the parameter combination ($ \alpha $, $ \beta $, $ \gamma $) = (10$ ^{-3} $, 2, 3). In these Hill surfaces, slopes of 1 correspond to free random (hyperbolic) binding, while non-unity slopes denote the existence of collective influences in the system. Specifically, steep slopes reflect a phenomenon of cooperativity increasing the sensitivity of the system to slight changes in ligand concentration. Near horizontal slopes and plateaus indicate the regions of relative TF inefficacy as long as the concentration of the other TF is limiting. This is a situation of negative cooperativity. These Hill landscapes highlight the differences between the basic (but doubtful) hierarchical mechanism (right panel of Fig.2b) and the nonequilibrium model (right panel of Fig.2c). While there is no limitation other than saturation in the response to large $ Y $ in the equilibrium model, this is no longer the case when $ Y>X $ in Fig.2c. In this respect, the latter model recovers some features of the independent system in which parallel increases of $ X $ and $ Y $ are necessary to allow their action. This property is related to the fact that $ A $ can always escape the promoter and is not trapped contrary to the equilibrium model. This difference could be used as a tool for experimentally probing hierarchical systems. The hierarchical nature of these systems is illustrated by the preponderant role of $ A $ at low fractional activity. Hence, the active chromatin remodeling mechanism described here allows pronounced non-linearity, even for monomeric TFs, which can be further enhanced by other modes of cooperativity.

\subsection{Cumulating the layers of cooperativity}

Hierarchical cooperativity provides an exquisite mode of sigmoidicity, in equilibrium (Fig.1a) as well as non-equilibrium conditions (Fig.1c,d). In the equilibrium system, joint sigmoidicity is obtained only in the $ A+B $ bisector, by intersecting two series of orthogonal hyperbolas (Fig.2b, illustrated by the 2D Hill curve at $ X=Y $). In addition, in the active remodeling model, the response to $ A $ alone is also sigmoidal (visible along the $ A $ axis in the 3D plot of Fig.2c). The self-cooperativity of $ A $ further enhances the steepness of the global response in the bisector ($ A+B $) (small 2D plot in Fig.2c). The maximal Hill coefficients ($  n_{H} $) for the different models are, for the independent TFs of Fig.2a (Eq.(11)):  $  n_{H}(A) = 1 $, $  n_{H}(B) = 1 $, $  n_{H}(A+B) = 1 $; for the equilibrium model of Fig.1a:  $  n_{H}(A) = 1 $, $  n_{H}(B) = 1 $, $  n_{H}(A+B) = 2 $ and for the non-equilibrium model of Fig.1d and Fig.2c: $  n_{H}(A) = 2 $, $  n_{H}(B) = 1 $, $  n_{H}(A+B) = 3 $. The sigmoidicity of this latter situation is illustrated in Fig.3. This source of sigmoidicity can surimpose to other ones, including: i) TF multimerization (neglected here) and ii) the cooperative recruitment of transcription machineries by DNA-bound TFs (Michel, 2010), rarely considered in transcription modeling studies. For simplicity, it has not been not taken into account in the present study and Eq.(1) describes additive contributions of the TFs $ A $ and $ B $ to the global promoter activity.

\section{Speculative role of the present mechanism for the MMTV}
Certain mouse strains contain without apparent trouble, genome-integrated MMTV which are vertically transmitted over generations. MMTV-infected mouse cells can also remain healthy. Though viruses are generally detrimental for infected cells, the issue of an infection for the host cells often depends on the conditions. As it is counter-productive for a stowaway to destroy his vehicle, viral infection is not necessarily lytic. Indeed, host genome-integrated viruses have the opportunity to propagate passively as furtive aliens, through the mere spreading of the host cells. Accordingly, they developed strategies during evolution to preserve host viability as long as living conditions are satisfactory. In turn, when the viability of the host cells is menaced, the lytic phase is triggered and leads to the production of metabolically inert viral particles which are more resistant to deleterious conditions. This strategy has been observed in the prokaryotic world, for example in the case of the lambda bacteriophage in lysogenic bacteria, but it can also apply to certain eukaryotic integration viruses, such as the MMTV provirus which generally remains dormant in adults, unless they are submitted to stresses. MMTV expression is weak in stressless conditions since nucleosomes ensure its transcriptional silencing. In cells containing GR, glucocorticoid hormones trigger MMTV expression. Glucocorticoid hormones (corticol, corticosterone), are the hormones of nervous stress, which activate the whole panoply of GR activities (nuclear import, DNA binding, transactivation, recruitement of BRG1). The secondary TFs which strongly enforce the GR action are NF1 but also Oct-1 or Oct-2. Interestingly, the preferential binding site for Oct-2 defined in (Rhee et al., 2001) precisely corresponds to the Oct module present in the MMTV promoter. It is inducible by bacterial lipopolysaccharides (LPS) and inflammatory signals. Hence, several types of stresses: nervous (GR) and infectious (Oct-2), concur to activate MMTV. The sigmoidal shape of the response shown in Fig.2c, is such that the combination of the two types of stress is required to trigger transcription. Moreover, the near horizontal slope of the transcription surface and the very low responsiveness to low GR and Oct concentrations (Fig.3), allow to buffer stochastic fluctuations of these TFs. By this way, MMTV can remain latent in moderately stressed cells, and is revived upon conjunction of stresses (Fig.4). 

\begin{center}
\includegraphics[width=7.5cm]{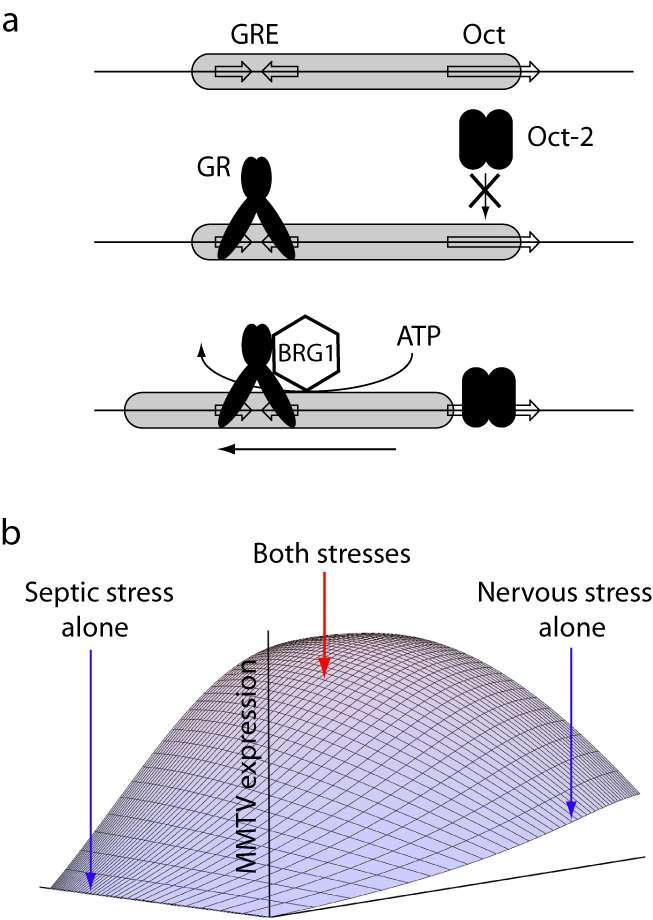} \\
\end{center}
\begin{small} \textbf{Figure 4.} Example of hierarchical transcriptional cooperativity mediated by chromatin remodelling. (\textbf{a}) Schematic representation of the proximal MMTV promoter critically regulated by a nucleosome (grey), which is positioned to overlap the binding sites for GR and Oct-2. In this configuration, only GR can bind its target site, owing to its capacity to interact with nucleosomal DNA. Its fixation then triggers the recruitment of the chromatin remodeller BRG1, which in turns allows the fixation of the Oct-2 factor requiring a fully accessible DNA helix.(\textbf{b}) This systems predicts an overadditive combination effect and the response of the MMTV to multiple stresses. \end{small}\\

This could be the case for example when the hosting mouse is both frightened, with production of glucocorticoid hormone (for example if a cat appears in the neighbourhood) and wounded (leading to a bacterial infection and to Oct-2 induction). When these conditions are reunited, the mouse's life is probabilistically compromised and it is beneficial for the MMTV to escape it before sinking with it. MMTV expression can be particularly important in lymphocytes because these cells are cellular reservoirs for MMTV, undergo apoptosis upon glucocorticoid exposure and display strong Oct-2 induction by inflammatory stress (Bendall et al., 1997). This transcriptional arbitration is equivalent to that of a crisis board but is more economic. The conversion of random interactions into discerning actions is a typical characteristic of dissipating systems, involving in the present case energy-dependent chromatin remodeling.

\section{Conclusion}
 
The model proposed here is a simplification omitting many actors in the MMTV promoter story, but is sufficient to reconcile conflicting data. While the first articles convincingly demonstrated that GR binding opens the way to NF1 and Oct-1, further studies showed that NF1 and Oct-1 are present prior to glucocorticoid hormone addition (Belikov et al., 2004). The presetting action of NF1 suggested in this latter article was interpreted as a locking action of NF1, that was suggested to click nucleosome positioning in a unique configuration. This interpretation is fully consistent with the mechanism proposed here, in which chromatin closing and NF1 binding are mutually exclusive events. In the present model, a fuzzy pattern is expected if the $ P $ and $ P' $ states and their transition intermediates coexist in the cell population. This coexistence is possible in presence of GR alone (ligand $ A $ in Fig.1c), but not of NF1 or Oct-1 (ligand $ B $ in Fig.1c). The $ P $ state corresponds to the positioning of nucleosomes thermodynamically favoured by nucleotide sequence-specific DNA bendability (Pina et al., 1990). The $ P' $ state is a less stable configuration, whose formation is forced by SWI/SNF ATPases and which is then locked by NF1/Oct-1 as long as present.
	This scheme is satisfactory in that it allows to explain previous observations seemingly contradictory: (i) the initiation role of activated GR on NF1/Oct-1 fixation, (ii) the presetting action of NF1/Oct-1 on GR exchanges, (iii) the fact that GR is not trapped in presence of NF1/Oct-1. Considering the abundance of chromatin remodeling factors in the cell (Rippe et al., 2007), such a mechanism could be very general and provide a widespread mode of cooperativity between TFs that do not directly interact with each other. In addition, these enzymes render cell-context specific, the role of ubiquitous actors such as the DNA-binding elements for TFs that are common to several cell types. Two nuclear receptors: GR and PR (progesterone receptor), are of equivalent strength and share the same DNA modules in the MMTV promoter; but interestingly, in a cellular context permissive for GR, PR fails to activate MMTV integrated into ordered chromatin, but induces MMTV when transfected in an open chromatin state (Smith et al., 1997). Accordingly, PR is unable to induce chromatin remodeling at stably integrated MMTV templates in these cells (Smith et al., 1997; Fryer and Archer, 1998) and the reciprocal situation is obtained in other cellular contexts (T47D, personal data).
The mechanism proposed in this study could be a pivotal device for the management of the eukaryotic genomes based on their nucleosomal organisation. It allows: (i) to solve apparent discrepancies between experimental observations, so far barely reconcilable in equilibrium conditions; (ii) to establish a primary and highly tunable mode of cooperativity between TFs, considering that the chromatin-remodeling enzymes are themselves subject to refined regulations; (iii) and to bypass the need for direct interactions between them, which is questioned in (Chu et al., 2009). \\

\textbf{Revision of the concept of pioneer transcription factors.}
Pioneer transcription factors are defined as developmental factors opening the way to secondary transcription factors. In this sequential mode of action, the pioneer transcription factor is envisioned as autonomous whereas the secondary transcription factor is tributary of the pioneer one. For example, the pioneer factors FOXA1 (also involved in the MMTV system), AP2$ \gamma $, PBX1 and GATA3, are supposed to preset chromatin and allow the fixation of the estrogen receptor-$ \alpha $ (ER$ \alpha $) in mammary lumenal epithelial cells. More than 80\% of the ER$ \alpha $-binding sites are associated to the fixation of one of these pioneer factors (Magnani and Lupien, 2014). The depletion of these factors prevents ER$ \alpha $ from binding, but the reverse has recently also been shown true (Caizzi et al., 2014), which singularly challenges the unilateral dependence of secondary transcription factors on pioneer factors. By contrast, the present model is compatible with a reciprocal dependence between these factors. As shown in Fig.1a,b, the secondary factor can strengthen the fixation of the pioneer factor by extending the fraction of time of the remodeled chromatin state, to which the pioneer factor can bind with a higher affinity. In addition, this mechanism is dynamic, contrary to a static hierarchical view FOXA1 $ \rightarrow  $ ER$ \alpha $, which is poorly compatible with the half residence time of FOXA1 of 4 minutes, as measured by fluorescence microscopy (Sekiya et al., 2009). The mechanism described here predicts mutual influences between transcription factors, creating a key combination effect for turning on or off gene expression.

\textit{The present manuscript is an extended version of the article: Hierarchical cooperativity mediated by chromatin remodeling; the model of the MMTV transcription regulation. Michel, D. 2011. J. Theor. Biol. 287, 74-81.}\\
 
Acknowledgement: Supported by the University of Rennes1, Action d\'efits scientifiques \'emergents.

\end{multicols}
\end{document}